\documentclass[aps,prl,superscriptaddress,amsmath,amssymb,floatfix,twocolumn]{revtex4-1}
\usepackage{graphicx}
\usepackage{subfigure}
\usepackage{color}
\usepackage{epstopdf}

\newcommand{\mbk}{\mathbf{k}}

\begin{document}

\title{
Orbital selectivity enhanced by nematic order in FeSe}

\author{Rong Yu}
\email{rong.yu@ruc.edu.cn}
\affiliation{Department of Physics and Beijing Key Laboratory of Opto-electronic Functional Materials and Micro-nano Devices, Renmin University of China, Beijing 100872, China}

\author{Jian-Xin Zhu}
\email{jxzhu@lanl.gov}
\affiliation{Theoretical Division and Center for Integrated Nanotechnologies, Los Alamos National Laboratory, Los Alamos, New Mexico 87545, USA}

\author{Qimiao Si}
\email{qmsi@rice.edu}
\affiliation{Department of Physics \& Astronomy, Rice University, Houston, Texas 77005,USA}
%\date{\today}

\begin{abstract}
Motivated by the recent low-temperature
experiments
on bulk FeSe, we study the electron correlation effects in a multiorbital model for this compound
in the nematic phase using the $U(1)$ slave-spin theory. We find that a finite nematic order helps to stabilize
an orbital selective Mott phase.
Moreover,
we propose that when
the $d$- and $s$-wave bond nematic orders are combined with
the ferro-orbital order, there exists a surprisingly
large orbital selectivity between the $xz$ and $yz$ orbitals
even though the associated band splitting
is relatively
 small. Our results
explain the
seemingly
unusual observation
of strong orbital selectivity in the nematic phase of FeSe,
%and
uncover new clues on the nature of the nematic order,
and
%sets
set the stage to elucidate
the interplay between superconductivity and nematicity
in
iron-based superconductors.
\end{abstract}

\maketitle

%\onecolumngrid

{\it Introduction.~}
The iron-based superconductors (FeSCs)
present a topic of extensive current research in condensed matter physics ~\cite{Kamihara2008,Johnston,Dai2015,NatRevMat:2016, Hirschfeld2016, FWang-science2011}.
One characteristic
 feature of these materials is that multiple electronic $3d$ orbitals are important for their
electronic structure. With the electron-electron
interactions
in these multiorbital systems,
the entwined degrees of freedom generate
a very rich phase diagram with a variety of correlation-induced electronic orders~\cite{Johnston,Dai2015,NatRevMat:2016, Hirschfeld2016, FWang-science2011}.

Besides the overall effect of electron correlations
~\cite{Basov.2009,Si2008,Haule08,Yi_PRL2015,Wang_PRB2015}, the multiple orbitals in the FeSCs may possess
 different degrees of
correlation effects.
Such an orbital selectivity has been found in multiorbital models for FeSCs~\cite{Yu2011,Yu_2013a,deMedici2014,Rincon_2014}.
%Due to
Because of the kinetic hybridization between the different orbitals in these models,
this effect is surprising and to be contrasted ~\cite{Yu_2017,Komijani_2017}
with what happens when the orbitals do not mix with each other
~\cite{Anisimov_2002,Koga_2004,WernerMillis_2007,Werner_2009,Medici_2009}.
It has been shown that the Hund's coupling
helps to stabilize an orbital-selective Mott phase (OSMP) inside which the Fe $3d_{xy}$ orbital is
Mott localized while the other orbitals are still itinerant~\cite{Yu_2013a}.
Many iron chalcogenides and pnictides
appear to be close to the OSMP in the phase diagram,
and can be driven into this phase by doping, applying pressure, or
% tuning
varying
temperature~\cite{Yi_Review2017,Yi_2013,Yi_NatComm2015,Pu_2016,Wang_2014,Ding_2014,Li_2014,Gao_2014,Das_2015}.

Another important aspect of the multiorbital effect in FeSCs is associated with the nematic order. In most of the undoped iron pnictides, there is a structural transition from a tetragonal phase to an orthorhombic one with lowering the temperature. Right at or slightly below the structural transition temperature, the system develops a long-range $(\pi,0)$ antiferromagnetic (AFM) order. The superconductivity usually appears near this antiferromagnetic phase. In between the structural and the magnetic transitions the $C_4$ lattice rotational symmetry is broken, and the system is in a nematic phase. The origin of this nematic phase is still under debate.
In the spin-driven-nematicity scenario, the nematicity is associated with an Ising order characterizing the anisotropic antiferromagnetic fluctuations~\cite{DaiSi_2009,FangKivelson:2008,XuMullerSachdev:2008,ChandraColemanLarkin:1990}
or the antiferroquadrupolar ones ~\cite{Yu_2015}.
The
corresponding bond nematicity
may
have different forms,
such as $d$- or $s$-wave nearest neighbor bond nematic orders~\cite{Su_2014}.
On symmetry grounds, this bond
nematicity is linearly coupled to a ferroorbital order that lifts the degeneracy of the Fe $d_{xz}$ and $d_{yz}$ orbitals.
Thus,
a ferroorbital order is also expected to be present.
Interestingly,
recent
angle-resolved photoemission spectroscopy (ARPES)
measurements on a variety of FeSCs observed a momentum dependent splitting between the $xz$- and $yz$-orbital dominant bands, which suggests the coexistence of several different nematic orders~\cite{Watson_2016,Zhang_2016,Zhang_2015}.

Among the FeSCs, FeSe is one of the most fascinating compounds. The single-layer FeSe on the SrTiO$_3$ substrate holds the record of the highest superconducting transition temperature of the FeSCs~\cite{FeSeSL}. On the other hand, the bulk FeSe has a structural transition at $T_s = 90$ K without showing an AFM long-range order down to the lowest accessible temperature under ambient pressure, suggesting an unusual magnetism in the ground state~\cite{Yu_2015}. In the nematic phase, ARPES measurements find a momentum dependent splitting between the $xz$- and $yz$-orbital dominant bands with
small splittings at both the $\Gamma$ and $M$ points of the Brilluion zone (BZ)~\cite{Watson_2016,Zhou_2018}.
Recent scanning tunneling microscopy (STM) experiments have revealed a strong orbital selectivity ~\cite{Davis_2017,Davis_2018}.
Especially, the estimated ratio of the quasiparticle weights between the $yz$ and $xz$ orbitals is very large: $Z_{yz}/Z_{xz}\sim 4$.
Because the band splittings are relatively small ~\cite{Watson_2016,Zhou_2018}, such a strong orbital selectivity
is surprising~\cite{Bascones_2017}.
It is important to resolve this puzzle, given that both the nematic correlations and orbital selectivity may be
of broad interest to unconventional superconductivity in the iron-based materials and beyond.

In this Letter, we examine the electron correlation effects in a multiorbital Hubbard model for the nematic phase of FeSe using previously developed $U(1)$ slave-spin theory~\cite{Yu_2012}. We consider three types of nematic orders, a ferro-orbital order, a $d$-wave nearest-neighbor bond order, and an $s$-wave nearest-neighbor bond order, and
analyze
their effects on the orbital selectivity. We solve the saddle-point equations and
show that the OSMP is promoted by
any of these nematic
orders. This effect is delicate,
because we also find that
the full Mott localization of the system depends on the type and strength of the nematic order.
Remarkably, we find that, by taking a proper combination of the three
types of
nematic order,
 the system can exhibit a strong orbital selectivity with $Z_{yz}/Z_{xz}\sim 4$ but rather small band splitting ($\lesssim 50$ meV)
 at the $\Gamma$ and $M$ points of the BZ.
 Our results naturally explain the
 unusually
  large orbital selectivity in the nematic phase of FeSe
  ~\cite{Davis_2017,Davis_2018},
thereby setting the stage to understand the superconducting state in this compound.
More generally, the necessity of coexisting nematic orders with comparable strength implies that the nematicity in the FeSCs cannot be entirely driven by the orbital order,
thereby providing new clues to the
origin of the nematicity in FeSCs.

{\it Model and method.~}
We study a five-orbital Hubbard model for FeSe. The Hamiltonian reads as
\begin{equation}\label{HamTot}
 H = H_{\rm{TB}} + H_{\rm{nem}} + H_{\rm{int}}.
\end{equation}
$H_{\rm{TB}}$ is a five-orbital tight-binding Hamiltonian with tetragonal lattice symmetry,
\begin{equation}
 \label{Eq:Ham_0} H_{\rm{TB}}=\frac{1}{2}\sum_{ij\alpha\beta\sigma} t^{\alpha\beta}_{ij}
 d^\dagger_{i\alpha\sigma} d_{j\beta\sigma} + \sum_{i\alpha\sigma} (\epsilon_\alpha-\mu) d^\dagger_{i\alpha\sigma} d_{i\alpha\sigma},
\end{equation}
where $d^\dagger_{i\alpha\sigma}$ creates an electron in orbital $\alpha$ ($\alpha=1,...,5$ denoting $xz$, $yz,$ $x^2-y^2$, $xy$, and $3z^2-r^2$ orbitals, respectively) with spin $\sigma$ at site $i$, $\epsilon_\alpha$
refers to the energy level associated with the crystal field splitting (which is diagonal in the orbital basis),
and $\mu$ is the chemical potential.
The tight-binding parameters $t^{\alpha\beta}_{ij}$ and $\epsilon_\alpha$,
which are presented in the Supplemental Material (SM)~\cite{SupplMat},
are determined by fitting to DFT band structure for FeSe,
and we
specify $\mu$ to fix the total electron density to $6$ per Fe. The on site interaction $H_{\rm{int}}$ reads
\begin{eqnarray}
 \label{Eq:Ham_int} H_{\rm{int}} &=& \frac{U}{2} \sum_{i,\alpha,\sigma}n_{i\alpha\sigma}n_{i\alpha\bar{\sigma}}\nonumber\\
 &&+\sum_{i,\alpha<\beta,\sigma} \left\{ U^\prime n_{i\alpha\sigma} n_{i\beta\bar{\sigma}}\right.
 + (U^\prime-J_{\rm{H}}) n_{i\alpha\sigma} n_{i\beta\sigma}\nonumber\\
&&\left.-J_{\rm{H}}(d^\dagger_{i\alpha\sigma}d_{i\alpha\bar{\sigma}} d^\dagger_{i\beta\bar{\sigma}}d_{i\beta\sigma}
 +d^\dagger_{i\alpha\sigma}d^\dagger_{i\alpha\bar{\sigma}}
 d_{i\beta\sigma}d_{i\beta\bar{\sigma}}) \right\}.
\end{eqnarray}
where $n_{i\alpha\sigma}=d^\dagger_{i\alpha\sigma} d_{i\alpha\sigma}$.
Here,
$U$, $U^\prime$, and $J_{\rm{H}}$, respectively denote the intraorbital repulsion, the interorbital repulsion,
and the Hund's rule coupling, and we take $U^\prime=U-2J_{\rm{H}}$.~\cite{Castellani78}
To study the model in the nematic phase, we introduce bare nematic orders in the $xz$ and $yz$ orbital subspace into $H_{\rm{nem}}$. In the momentum space
\begin{eqnarray}\label{Eq:Ham_nem}
 && H_{\rm{nem}} = \sum_\mbk \left[ -2\delta_d (\cos k_x-\cos k_y)(n_{k1}+n_{k2}) \right.\nonumber\\
 && - \left.  2\delta_s (\cos k_x + \cos k_y) (n_{k1}-n_{k2}) + \delta_f (n_{k1}-n_{k2}) \right].
\end{eqnarray}
Here,besides the ferro-orbital order ($\delta_f$) we have also taken into account a $d$- and an $s$-wave bond nematic order ($\delta_d$ and $\delta_s$), which corresponds to nearest-neighboring hopping anisotropy.~\cite{Su_2014}

We investigate
the electron correlation effects by using a $U(1)$ slave-spin theory~\cite{Yu_2012}. In this approach, we rewrite $d^\dagger_{i\alpha\sigma} = S^+_{i\alpha\sigma} f^\dagger_{i\alpha\sigma}$, where $S^+_{i\alpha\sigma}$ ($f^\dagger_{i\alpha\sigma}$) is the introduced quantum $S=1/2$ spin (fermionic spinon) operator to carry the charge (spin) degree of freedom of the electron at each site. For a general multiorbital model three saddle-point solutions can be stabilized: a metallic state with the quasiparticle spectral weight $Z_{\alpha}>0$ in all orbitals, a Mott insulator with $Z_{\alpha}=0$ in all orbitals, and an OSMP with $Z_{\alpha}=0$ in some orbitals but $Z_{\alpha}>0$ in other orbitals. In the metallic state, a significant effect of the electron correlations is that the electron band structure is renormalized by $Z_\alpha$ and the effective on site potential $\tilde{\mu}_\alpha$.~\cite{SupplMat} We are particularly interested in how the band splittings between the $xz$- and $yz$-dominant bands at the $\Gamma$ and $M$ points of the BZ ($\Delta E_\Gamma$ and $\Delta E_{\rm{M}}$) evolves with interaction $U$ and nematic order $\delta_a$ ($a=f$, $d$, $s$).
Keeping in mind the aim of understanding the effect of nematicity on the orbital selectivity,
we simplify our
analysis by focusing on the diagonal part of $J_{\rm{H}}$
(SM, end of the 2nd section~\cite{SupplMat}).

\begin{figure}[t!]
\centering\includegraphics[%scale=0.28
width=75mm
]{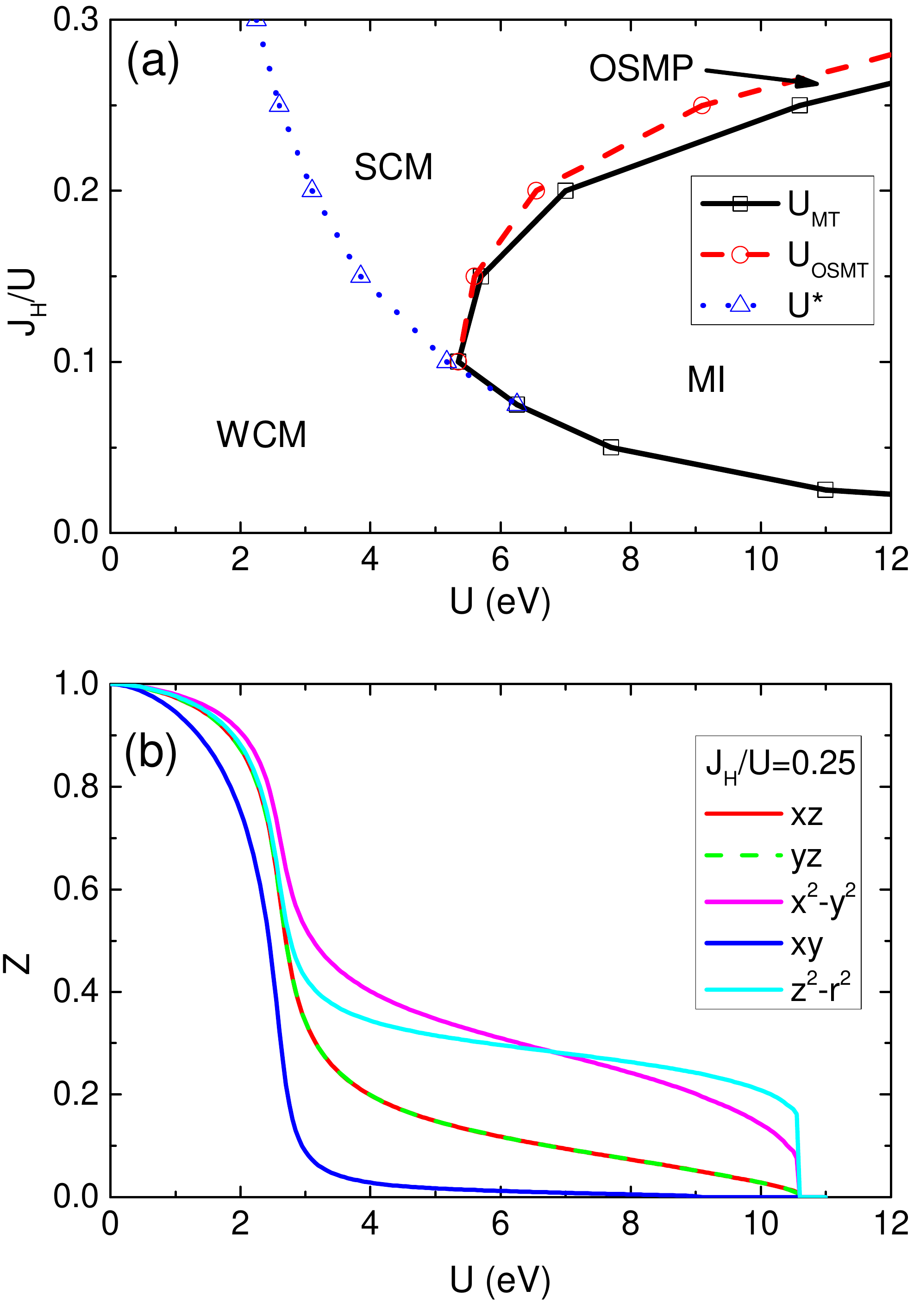}
\caption{(Color online) (a): Ground-state phase diagram of the five-orbital Hubbard model for FeSe in the tetragonal phase. (b): Evolution of the orbital resolved quasiparticle spectral weights with increasing $U$ at $J_{\rm{H}}/U=0.25$.
}
\label{fig:1}
\end{figure}

{\it Phase diagram in the tetragonal phase.~} We first
examine
the correlation effects in the tetragonal phase. The ground-state phase diagram in the $J_{\rm{H}}$-$U$ plane is shown in Fig.~\ref{fig:1}(a). It contains three phases: a metal, a MI, stabilized for $U\gtrsim5$ eV, and an OSMP close to the boundary of the MI when $J_{\rm{H}}/U\gtrsim0.1$. In the OSMP, the $xy$ orbital is Mott localized while other Fe $3d$ orbitals are still itinerant (Fig.~\ref{fig:1}(b)). In the metallic phase, there is a crossover at $U^\star$ between a weakly correlated metal (WCM) and a strongly correlated metal (SCM). $Z_{\alpha}$ drops rapidly with increasing $U$ across $U^\star$ (Fig.~\ref{fig:1}(b)). Qualitatively, the phase diagram here for FeSe is similar to those for other iron chalcogenides~\cite{Yi_NatComm2015,Yu_2013a}. By comparing with ARPES results on FeSe$_x$Te$_{1-x}$~\cite{Liu_2015}, it is extrapolated that $J_{\rm{H}}/U\sim0.15$-$0.3$ eV, and $U\sim2.5$-$4$ eV in FeSe, suggesting that FeSe is close to the crossover line $U^\star$ in the phase diagram, and has moderate orbital selectivity compared to FeTe~\cite{Liu_2015}. However, the tetragonal phase of FeSe is only stabilized above the structural transition. As shown in
Fig.S3 of the
%Supplemental Material,
SM~\cite{SupplMat},
the
threshold $U$ value for the orbital-selective Mott transition (OSMT) decreases with increasing temperature.
Thus,
 for $T\gtrsim90$ K in the tetragonal phase, the system may already be close to the boundary of the OSMP.

\begin{figure}[t]
\centering\includegraphics[%scale=0.28
width=75mm]{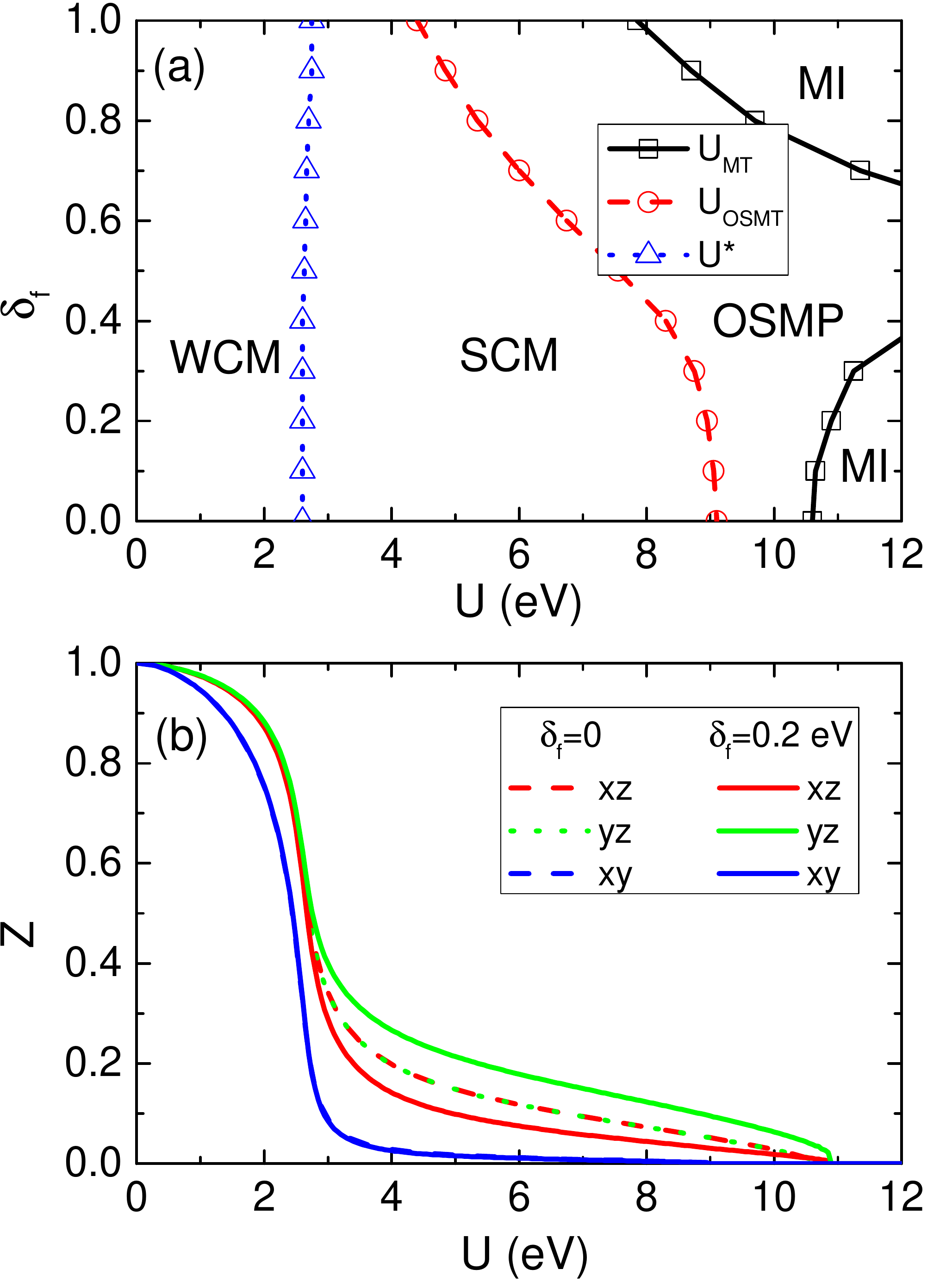}
\caption{(Color online)
(a): Ground-state phase diagram of the five-orbital Hubbard model for FeSe with a ferro-orbital order $\delta_f$ at $J_{\rm{H}}/U=0.25$. (b): The quasiparticle spectral weights in the $t_{2g}$ orbital sector with and without a ferro-orbital order.
}
\label{fig:2}
\end{figure}
{\it Enhanced orbital selectivity in the nematic phase.~}
We turn next to how the nematicity influences electron correlations. Fig.~\ref{fig:2}(a) shows how the phase diagram varies with the bare ferro-orbital order $\delta_f$ at $J_{\rm{H}}/U=0.25$. The phase boundaries change very little for $\delta_f\lesssim0.2$ eV (see also Fig.~\ref{fig:2}(b)).
Further increasing $\delta_f$,
$U^\star$ slightly increases. This is because $U^\star$ corresponds to an energy scale for the overall correlation effect, where a high-spin $S\sim2$ state is approximately formed.\cite{Yu_2013a} By increasing $\delta_f$, the $d_{xz}$ and $d_{yz}$ orbitals are driven away from half-filling. Therefore, a larger $U$ value is needed to push these orbitals back to
being
close to half-filling to form the high-spin state.
On the other hand, the critical $U$ for the OSMT significantly decreases,
 indicating an enhancement of orbital selectivity by the nematic order. This can be understood as follow:
 For a small $\delta_f$, the electron densities at $U=0$ in all three $t_{2g}$ orbitals are close to half-filling
(Fig.S4).
 But for a large $\delta_f$, since it lifts the $xz/yz$-orbital degeneracy, the electron densities $n_{xz}$ and $n_{yz}$ are highly different and away from half-filling, but $n_{xy}$ still
 is
  close to half-filling at $U=0$
 (Fig.S5).
 This makes the Mott localization of the $xy$ orbital much easier for large $\delta_f$. However, the critical $U$ for the full Mott localization first increases with $\delta_f$ then decreases for $\delta_f\gtrsim0.5$ eV. For small $\delta_f$, the $xz/yz$ orbitals are nearly degenerate, and a splitting between them effectively increases the total bare bandwidth, making the Mott localization of all orbitals harder. But further increases $\delta_f$, the center of the $yz$ band is shifted much lower than the other four. With a moderate $U$, it can be driven to a band insulator. Once this takes place, the other bands would be at half-filling, which is known to be the easiest to be Mott localized than at any other commensurate filling.

We also analyze
the effects of the two bond nematic orders on the Mott localization, and find that the enhancement of orbital selectivity is a general feature
(Fig.S6),
but a MI is disfavored. In the tight-binding model for FeSe
~\cite{SupplMat},
the nearest-neighbor hoppings along the $\hat{x}$ and $\hat{y}$ directions within the $xz$ orbitals (also within the $yz$ orbitals), $t^{11}_{\hat{x}(\hat{y})}$ (and $t^{22}_{\hat{x}(\hat{y})}$) are highly anisotropic. In particular,
$t^{11}_{\hat{x}}=t^{22}_{\hat{y}}\approx0$.
Hence either a $d$- or an $s$-wave bond nematic order will enhance the kinetic energy associated with the $xz$ and $yz$ orbitals. This increases the orbital selectivity, promoting an OSMP. But the overall bandwidth is also increased, and therefore destabilzes a MI.

\begin{figure}[t!]
\centering\includegraphics[%scale=0.28
width=75mm
]{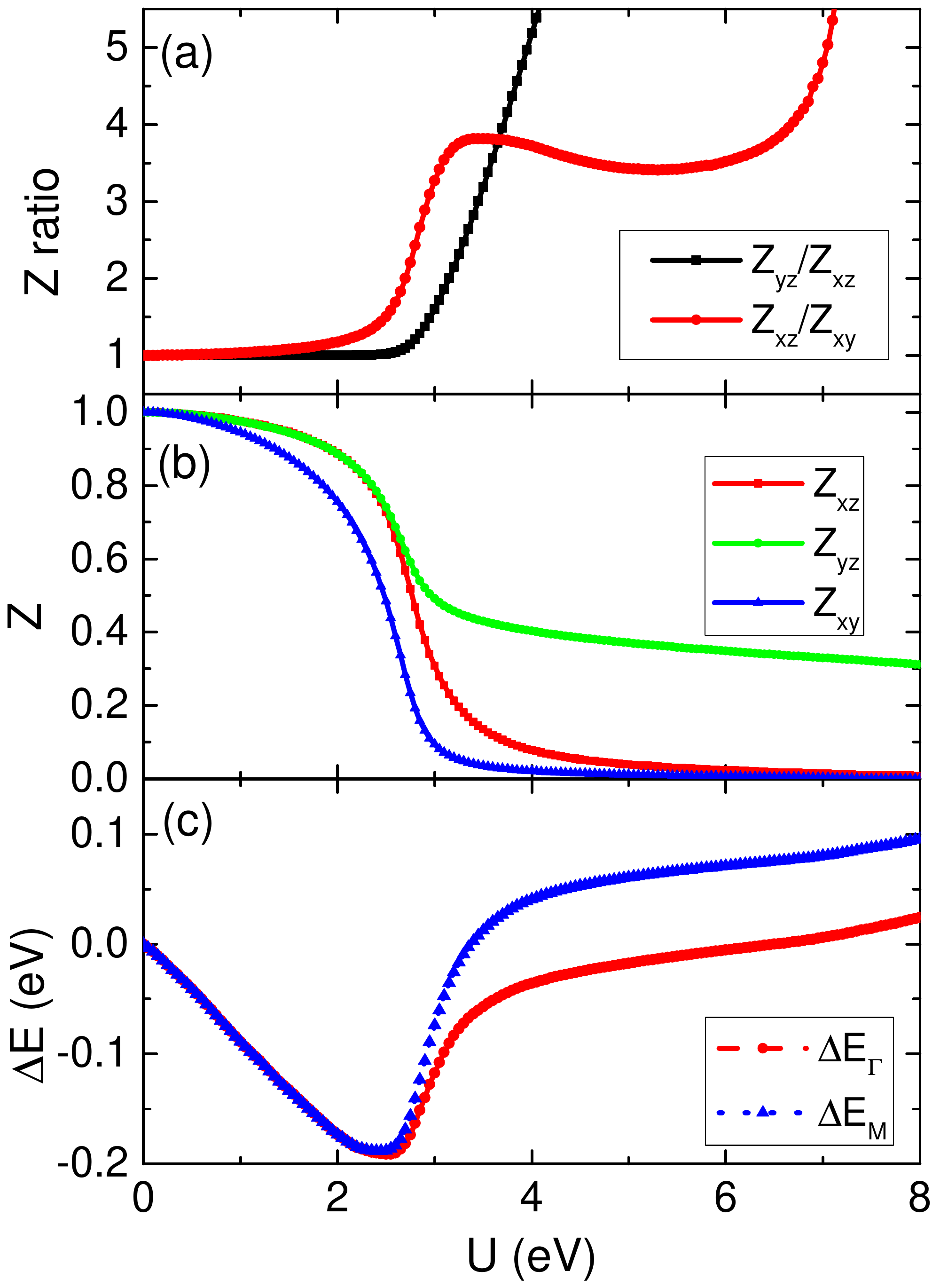}
\caption{(Color online) The orbital selectivity and band splitting in the nematic phase with a combined nematic order $\delta_f/4=\delta_d=\delta_s=0.2$ eV and with $J_{\rm{H}}/U=0.25$. (a): $Z_{yz}/Z_{xz}$ and $Z_{xz}/Z_{xy}$; (b): $Z_{xz}$, $Z_{yz}$, and $Z_{xy}$; (c): band splitting at $\Gamma$ ($\Delta E_\Gamma$) and M ($\Delta E_{\rm{M}}$) of the 2-Fe BZ.
}
\label{fig:3}
\end{figure}
{\it Orbital selectivity and band splitting.~} The nematic order not only helps stabilizing an OSMP by Mott localizing the $xy$ orbital, but also enhances the orbital selectivity between the $xz$ and $yz$ orbitals. As shown in Fig.~\ref{fig:2}(b), the $xz$ orbital is more correlated than in the tetragonal phase,
while the $yz$ orbital is less
so. The ratio $Z_{yz}/Z_{xz}$ increases with $\delta_f$ monotonically.
As mentioned earlier,
 recent STM experiments have observed $Z_{yz}/Z_{xz}\sim4$ in the nematic phase of FeSe~\cite{Davis_2017,Davis_2018}.

In the case of a single bare ferro-orbital order,
$\delta_f$ must be larger than $0.4$ eV
to arrive at such a large ratio within a reasonable range of $U$
(See Fig.S7).
This leads to the band splittings $\Delta E_\Gamma$ and $\Delta E_{\rm{M}}$ higher than 100 meV, which is much larger
 than the observed values ($\lesssim50$ meV).~\cite{Watson_2015,Watson_2016,Zhang_2016,Zhang_2015}
 Similar issue applies to the bond nematic orders alone (Fig.S7).
Thus, it is seemingly impossible to reconcile the contrasting properties as observed in STM and ARPES, respectively.

To make progress, we consider a combination of the three nematic orders.
An observation of Eq.~\eqref{Eq:Ham_nem} is that the bare band splittings at both $\Gamma$ and M points will be exactly canceled when taking $\delta_f/4=\delta_d=\delta_s$ (see
SM~\cite{SupplMat}). For definiteness,
we simply take this combined nematic order.
As shown in Fig.~\ref{fig:3}, for $U\sim3.5$-$4$ eV, such a combined nematic order gives $Z_{yz}\approx0.5$, $Z_{xz}\approx0.15$, and $Z_{xy}\approx0.05$, close to the experimentally determined values. Moreover, though the electron correlations renormalize the band splittings,
the cancellation effect is still prominent:
The band splittings $\Delta E_\Gamma$ and $\Delta E_{\rm{M}}$ are less than $50$ meV; this result is fully consistent
with the ARPES results.

\begin{figure}[t!]
\centering\includegraphics[%scale=0.28
width=75mm
]{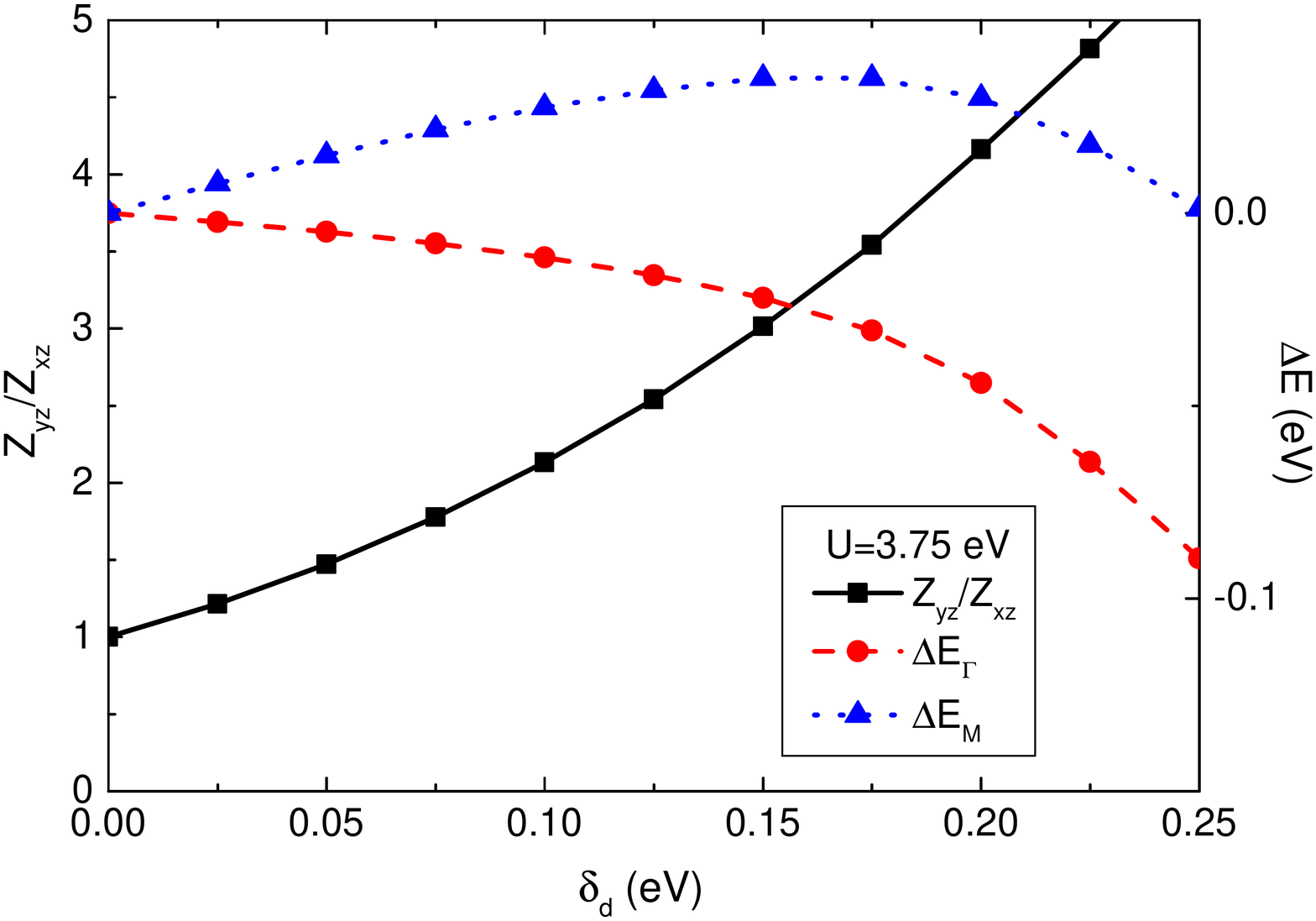}
\caption{(Color online) Evolution of $Z_{yz}/Z_{xz}$ (black solid) and band splittings at $\Gamma$ and M points, $\Delta E_\Gamma$ (red dashed) and $\Delta E_{\rm{M}}$ (blue dot), with the combined nematic order $\delta_f/4=\delta_d=\delta_s$ at $J_{\rm{H}}/U=0.25$ and $U=3.75$ eV.
}
\label{fig:4}
\end{figure}

We further show how the orbital selectivity and band splitting evolve with this combined nematic order
in Fig.~\ref{fig:4}. We find it quite remarkable
that a large orbital selectivity ($Z_{yz}/Z_{xz}$) while,
at the same time, a small band splitting
is stabilized by a
moderate
 (bare) combined nematic order.

{\it Discussions.~} With a single nematic order alone, to keep the band splittings $\Delta E_{\Gamma}$ and $\Delta E_{\rm{M}}$ to be compatible to the observed values ($< 50$ meV), we find that the bare nematic order must be small, leading to a weak orbital selectivity in the $xz$ and $yz$ sector with $Z_{yz}/Z_{xz}$ close to $1$. This is consistent with a previous study~\cite{Bascones_2017}.
Our calculations, however, have demonstrated a new effect:
with a proper combination of the bond nematic orders and the ferro-orbital order, the band splittings at the $\Gamma$ (and $M$) point caused by the different nematic orders compensate. In this way, moderate bare nematic orders can give rise to a strong orbital selectivity with $Z_{yz}/Z_{xz}\sim 4$, while keeping the band splittings near the Fermi level small at both the $\Gamma$ and $M$ points of the BZ.
This result
is robust against nematic quantum fluctuations, given that the system is not close to a nematic quantum critical point even though the band splittings are small.
 The large orbital selectivity manifests the effect of electron correlations. The latter is also implicated by the fact that
 the anisotropy in the optical conductivity induced by the nematic order extends to a large energy range, all the way to
%at least
about
 $0.5$ eV ({\it i.e.}, about 50 times of $k_B T_s$.)
~\cite{Chinotti_2017}.

The necessity that all the three types of nematic orders coexist implies that the nematic order observed in FeSCs has an unconventional origin
and cannot be entirely driven by orbital order.
 In the
spin driven nematicity, the nematic order is just an Ising order associated with short-range antiferromagnetic or antiferroquadrupolar orders within an effective
frustrated spin model including short-range Heisenberg and biquadratice interactions. Within this scenario, it is expected that the nearest-neighbor bond nematic orders,
together with the linearly coupled ferro-orbital order, contribute significantly in the nematic phase~\cite{Daghofer}, leading to a combined nematic order.
Our results thus suggest
that
 the nematicity in FeSCs
likely has a magnetic origin.

{\it Conclusions.~}
We have studied the effects of electron correlation with a nematic order in a multiorbital Hubbard model for FeSe by using the slave-spin method. We show that the orbital selectivity is generally enhanced by the nematic order.
A large combined nematic order can give rise to a large orbital selectivity in the $xz$/$yz$ orbital subspace with a small band splitting.
Our results
resolve an outstanding puzzle in the recent experimental observations
on the orbital selectivity and nematicity in FeSe,
elucidate the nature and
origin of the nematic order in FeSCs, and
pave
 the way for understanding the interplay between nematicity and high temperature superconductivity.

%\acknowledgements
\begin{acknowledgments}
We thank E. Abrahams, B. M. Andersen, E. Bascones, M. Daghofer, P. C. Dai, H. Hu, D. H. Lu, M. Yi, and X.-J. Zhou for useful discussions. This work has in part been supported by
the National Science Foundation of China Grant No. 11674392, Ministry of Science and Technology of China, National Program on Key Research Project Grant No.2016YFA0300504 and Research Funds of Remnin University
of China Grant No. 18XNLG24 (R. Y.),
and by
the U.S. Department of Energy, Office of Science, Basic Energy Sciences, under Award No. DE-SC0018197,
the Robert A.\ Welch Foundation Grant No.\ C-1411
and
a QuantEmX grant from ICAM and the Gordon and Betty Moore Foundation through Grant No. GBMF5305 (Q.S.), and by the U.S. DOE Office of Basic Energy Sciences E3B5 (J.-X.Z.).
The work was supported in part by the Center for Integrated Nanotechnologies, a U.S. DOE BES user facility.
Q.S. acknowledges the hospitality of University of California at Berkeley
and of the Aspen Center for Physics (NSF grant No. PHY-1607611).
\end{acknowledgments}

%%%%%%%%%%%%%%%%%%%%%%%%%%%%%%%%%%
%%%  Supplementary Materials   %%%
%%%%%%%%%%%%%%%%%%%%%%%%%%%%%%%%%%

\newpage
\setcounter{figure}{0}
\makeatletter
\renewcommand{\thefigure}{S\@arabic\c@figure}
\onecolumngrid
\section{ SUPPLEMENTAL MATERIAL -- Orbital selectivity enhanced by nematic order in FeSe}
%\newpage

%\onecolumngrid

%\section*{{\Large Supplementary Material for EPAPS}}

\subsection{Details on the Tight-binding parameters}
\begin{figure}[h!]
\centering\includegraphics[%scale=0.28
width=100mm
]{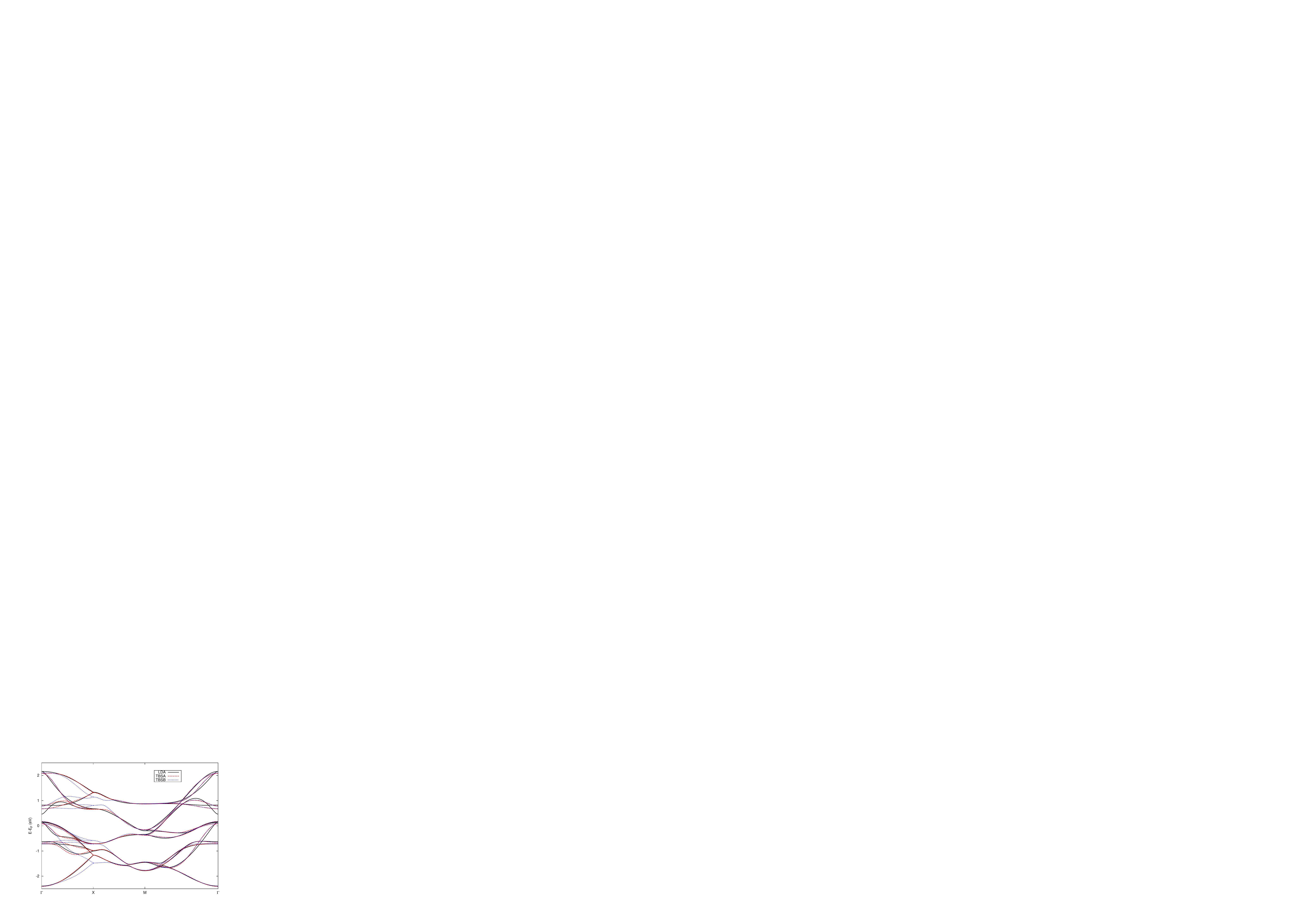}
\caption{(Color online)
Comparison of bandstructures of FeSe in the tetragonal phase from the LDA and tight-binding models. TBSA and TBSB denote tight-binding models with parameter set A and set B, respectively.}
\label{Sfig:1}
\end{figure}

\begin{figure}[h!]
\centering\includegraphics[%scale=0.28
width=100mm
]{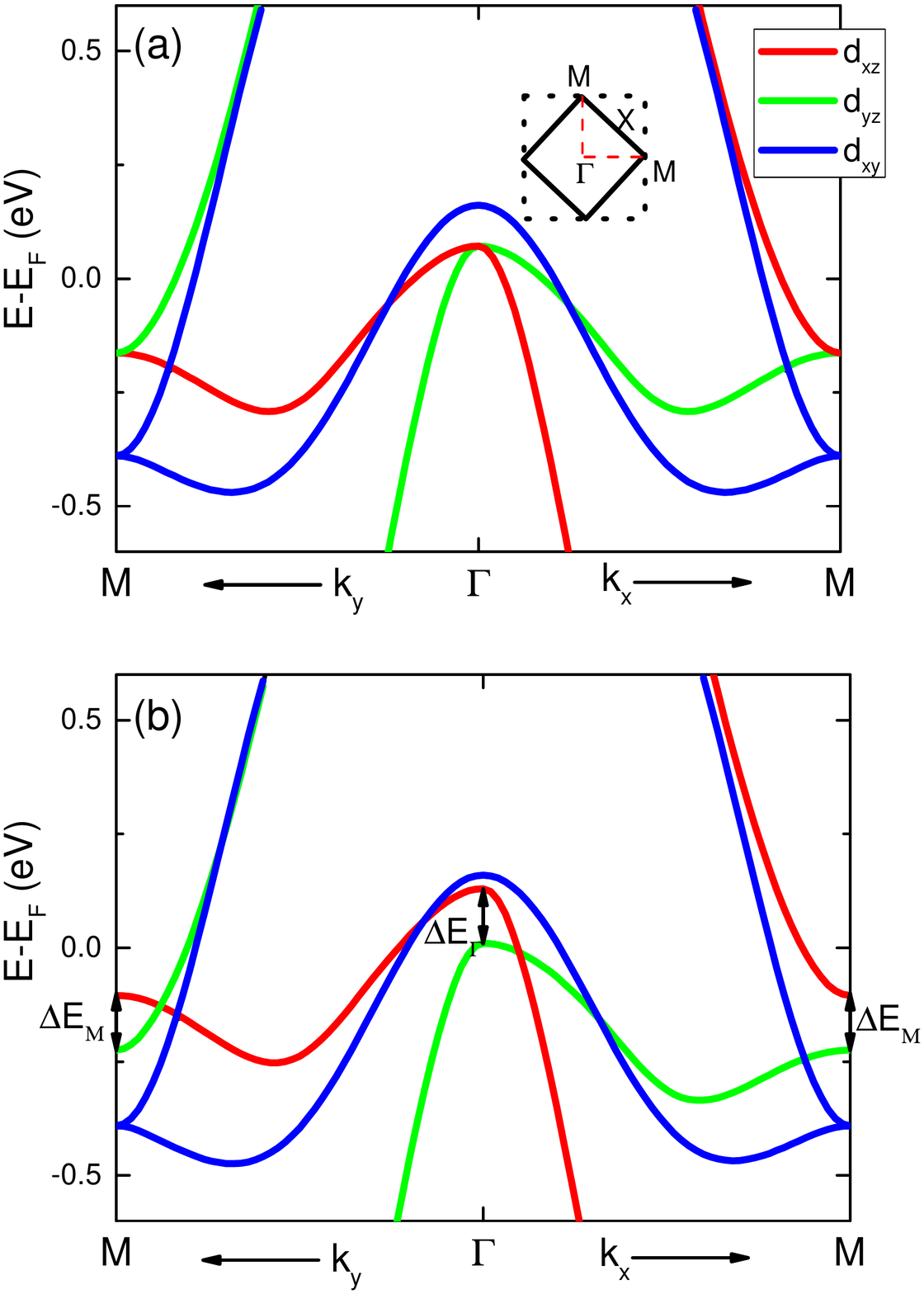}
\caption{(Color online)
(a): Bandstructure of the tight-binding model (with parameter set A) for FeSe in the tetragonal phase.
Different colors refer to dominant orbital characters of the relevant bands.
The $d_{xz}$ and $d_{yz}$ orbitals are degenerate at $\Gamma$ and M
points due to the tetragonal symmetry. The Inset shows the 1-Fe (black dashed) and 2-Fe (solid) Brillouion zones.
(b): Same as in (a) but in the nematic phase with a ferro-orbital order
to illustrate the band splittings at the $\Gamma$ ($\Delta E_\Gamma$) and M ($\Delta E_{\rm{M}}$) points,
which reflect the
 breaking
 of the
 tetragonal symmetry.}
\label{Sfig:2}
\end{figure}

To obtain the tight-binding parameters, we perform local density approximation (LDA) calculations for bulk FeSe
with a tetragonal structure, and we fit the LDA band structure to the tight-binding Hamiltonian.
The form of the five-orbital tight-binding Hamiltonian given in Ref.~[48] %\onlinecite{SM_Graser_2009}]
is used. We obtain two sets of tight-binding parameters, either of which well reproduces the LDA bandstructure.
The
two sets of tight-binding parameters  are listed in Table S1 and Table S2, respectively.
A comparison between the
 bandstructure from the LDA
 and those from the tight-binding model is shown in Fig.~\ref{Sfig:1}.
For these two sets of parameters, we obtain similar results in the slave-spin calculations in both the tetragonal and nematic phases.
For definiteness, we only present results with parameter set A.
In Fig.~\ref{Sfig:2}, we
illustrate
 the bands relevant to the bandsplittings $\Delta E_\Gamma$ and $\Delta E_{\rm{M}}$ in the nematic phase.

%\newpage
%\begin{table}[h]
\begin{minipage}{\linewidth}
%  \centering
\begin{center}
\begin{tabular}{cccccccc}
  \hline
  \hline
    & $\alpha=1$ & $\alpha=2$ & $\alpha=3$ & $\alpha=4$ & $\alpha=5$ &   &   \\ \hline
  $\epsilon_\alpha$ & -0.00733 & -0.00733 & -0.52154 & 0.10974 & -0.5694 &  & \\ \hline\hline
  $t^{\alpha\alpha}_\mu$ & $\mu=x$ & $\mu=y$ & $\mu=xy$ & $\mu=xx$ & $\mu=xxy$ & $\mu=xyy$ & $\mu=xxyy$ \\ \hline
  $\alpha=1$ & -0.0111 & -0.49155 & -0.23486 & -0.0119 & -0.04025 & -0.03917 & -0.03808\\ \hline
  $\alpha=3$ & -0.38485 &  & -0.08015 & -0.00646 &  &  &  \\ \hline
  $\alpha=4$ & -0.16872 &  & -0.10728 & -0.00626 & -0.04592 &  & -0.02079 \\ \hline
  $\alpha=5$ & -0.03681 &  &  & -0.00159 & -0.01585 &  & -0.02739 \\ \hline\hline
  $t^{\alpha\beta}_\mu$ & $\mu=x$ & $\mu=xy$ & $\mu=xxy$ & $\mu=xxyy$ &  &  &  \\ \hline
  $\alpha\beta=12$ &  & -0.12701 & -0.00655 & -0.05869 &  &  &  \\ \hline
  $\alpha\beta=13$ & -0.36123 & -0.07201 & -0.0134 &  &  &  &  \\ \hline
  $\alpha\beta=14$ & -0.20068 & -0.03548 & -0.00705 &  &  &  &  \\ \hline
  $\alpha\beta=15$ & -0.08057 & -0.14823 &  & -0.01218 &  &  &  \\ \hline
  $\alpha\beta=34$ &  &  & -0.0217 &  &  &  &  \\ \hline
  $\alpha\beta=35$ & -0.29868 &  & -0.01332 &  &  &  &  \\ \hline
  $\alpha\beta=45$ &  & -0.13208 &  & -0.05213 &  &  &  \\ \hline
  \hline
\end{tabular}
\par
\end{center}
\bigskip
\noindent
{\bf Supplemental Table S}1.
 Tight-binding parameter set A of the five-orbital model for
 bulk FeSe with the tetragonal structure.
Here we
use the same notation as in Ref.~
[48]. %\onlinecite{SM_Graser_2009}].
The orbital index $\alpha=$1,2,3,4,5 correspond to $d_{xz}$, $d_{yz}$,
$d_{x^2-y^2}$, $d_{xy}$, and $d_{3z^2-r^2}$ orbitals, respectively.
%The units of the parameters are eV.
The listed parameters are in eV.
\end{minipage}

%\newpage
%\begin{table}[h]
\begin{minipage}{\linewidth}
%  \centering
\begin{center}
\begin{tabular}{cccccccc}
  \hline
  \hline
    & $\alpha=1$ & $\alpha=2$ & $\alpha=3$ & $\alpha=4$ & $\alpha=5$ &   &   \\ \hline
  $\epsilon_\alpha$ & -0.04462 & -0.04462 & -0.46482 & 0.05510 & -0.48664 &  & \\ \hline\hline
  $t^{\alpha\alpha}_\mu$ & $\mu=x$ & $\mu=y$ & $\mu=xy$ & $\mu=xx$ & $\mu=xxy$ & $\mu=xyy$ & $\mu=xxyy$ \\ \hline
  $\alpha=1$ & -0.02391 & -0.46407 & -0.23723 & -0.01977 & -0.02735 & -0.03390 & -0.04816\\ \hline
  $\alpha=3$ & -0.38498 &  & -0.08201 & -0.01662 &  &  &  \\ \hline
  $\alpha=4$ & -0.14137 &  & -0.1046 & -0.02346 & -0.03301 &  & -0.02421 \\ \hline
  $\alpha=5$ & -0.02272 &  &  & -0.00983 & -0.00799 &  & -0.04141 \\ \hline\hline
  $t^{\alpha\beta}_\mu$ & $\mu=x$ & $\mu=xy$ & $\mu=xxy$ & $\mu=xxyy$ &  &  &  \\ \hline
  $\alpha\beta=12$ &  & -0.10648 & -0.00963 & -0.06207 &  &  &  \\ \hline
  $\alpha\beta=13$ & -0.33982 & -0.09231 & -0.03202 &  &  &  &  \\ \hline
  $\alpha\beta=14$ & -0.27634 & -0.0472 & -0.00575 &  &  &  &  \\ \hline
  $\alpha\beta=15$ & -0.08485 & -0.13135 &  & -0.0131 &  &  &  \\ \hline
  $\alpha\beta=34$ &  &  & -0.03011 &  &  &  &  \\ \hline
  $\alpha\beta=35$ & -0.34386 &  & -0.01006 &  &  &  &  \\ \hline
  $\alpha\beta=45$ &  & -0.0591 &  & -0.02495 &  &  &  \\ \hline
  \hline
\end{tabular}
\par
\end{center}
\bigskip
\noindent
{\bf Supplemental Table S}2.
 Tight-binding parameter set B of the five-orbital model for
 bulk FeSe with the tetragonal structure.
\end{minipage}

\subsection{Details on the $U(1)$ slave spin theory}
In this section, we
% present a summary
%on of
summarize
 the $U(1)$ slave-spin theory.
For further details, we refer to Refs.~[10] %\onlinecite{SM_Yu_2012}]
and [16].%\onlinecite{SM_Yu_2017}].

In the $U(1)$ slave-spin formulation, we introduce a quantum $S=1/2$ spin operator and use its XY component ($S^+_{i\alpha\sigma}$) to represent
the charge degree of freedom of the electron at each site $i$, in each orbital $\alpha$
and for each spin flavor $\sigma$. Correspondingly, we use a fermionic ``spinon'' operator
($f^\dagger_{i\alpha\sigma}$) to carry the spin degree of freedom.
The electron creation operator is
%then
represented as follows,
\begin{equation}\tag{S1}
 \label{Eq:SSCreate}
 d^\dagger_{i\alpha\sigma} = S^+_{i\alpha\sigma} f^\dagger_{i\alpha\sigma}.
\end{equation}
This
representation has an enlarged Hilbert space
compared to
 the
 %original
 one for the physical $d$ electrons. To
 restrict
 the Hilbert space to the physical one, we implement a local constraint,
\begin{equation}\tag{S2}
 \label{Eq:constraint}
 S^z_{i\alpha\sigma} = f^\dagger_{i\alpha\sigma} f_{i\alpha\sigma} - \frac{1}{2}.
\end{equation}

This representation contains
a $U(1)$ gauge redundancy corresponding to
$f^\dagger_{i\alpha\sigma}\rightarrow f^\dagger_{i\alpha\sigma} e^{-i\theta_{i\alpha\sigma}}$
and $S^+_{i\alpha\sigma}\rightarrow S^+_{i\alpha\sigma} e^{i\theta_{i\alpha\sigma}}$.
As a result,
 the slave spins can be used to
 carry the
 $U(1)$-symmetric
 physical charge degree of freedom,
 similarly as in the slave-rotor approach [49]. %~\cite{SM_FlorensGeorges}.

To ensure that the saddle point captures
the correct
quasiparticle spectral weight in the non-interacting
limit (being equal to $1$), we define a dressed operator in the Schwinger
boson representation of the slave spins (in a way similar to the standard
slave-boson theory~[50]):%\cite{SM_KotliarRuckenstein}):
\begin{equation}\tag{S3}
 \label{Eq:Zdagger}
 \hat{z}^\dagger_{i\alpha\sigma} = P^+_{i\alpha\sigma} a^\dagger_{i\alpha\sigma} b_{i\alpha\sigma}
 P^-_{i\alpha\sigma},
\end{equation}
where $P^\pm_{i\alpha\sigma}=1/\sqrt{1/2+\delta \pm (a^\dagger_{i\alpha\sigma} a_{i\alpha\sigma}
- b^\dagger_{i\alpha\sigma} b_{i\alpha\sigma})/2}$, and $\delta$ is an infinitesimal positive
number to regulate $P^\pm_{i\alpha\sigma}$.

Here $a_{i\alpha\sigma}$ and $b_{i\alpha\sigma}$ are Schwinger bosons representing the slave-spin operators:
$S^+_{i\alpha\sigma} = a^\dagger_{i\alpha\sigma} b_{i\alpha\sigma}$, $S^-_{i\alpha\sigma}
= b^\dagger_{i\alpha\sigma} a_{i\alpha\sigma}$, and
$S^z_{i\alpha\sigma} = (a^\dagger_{i\alpha\sigma}
a_{i\alpha\sigma} - b^\dagger_{i\alpha\sigma} b_{i\alpha\sigma})/2$.
They satisfy an additional constraint,
\begin{equation}\tag{S4}
\label{Eq:hard-core}
a^\dagger_{i\alpha\sigma} a_{i\alpha\sigma} + b^\dagger_{i\alpha\sigma} b_{i\alpha\sigma} = 1 .
\end{equation}
In other words, they are hard-core bosons.
In this representation, the constraint in Eq.~\eqref{Eq:constraint}
 becomes
\begin{equation}\tag{S5}
\label{Eq:constraint-in-bosons}
 a^\dagger_{i\alpha\sigma} a_{i\alpha\sigma} - b^\dagger_{i\alpha\sigma} b_{i\alpha\sigma}= 2 f^\dagger_{i\alpha\sigma} f_{i\alpha\sigma} -1 .
 \end{equation}
In addition,
Eq.~\eqref{Eq:SSCreate} becomes
\begin{equation}\tag{S6}\label{Eq:SBcreate}
d^\dagger_{i\alpha\sigma}=\hat{z}^\dagger_{i\alpha\sigma} f^\dagger_{i\alpha\sigma}.
\end{equation}

The Hamiltonian given in Eq. (1) of the main text
can then be effectively
 rewritten as
%\begin{eqnarray}
\begin{align}\tag{S7}
\label{Eq:HamSS}
H &= \frac{1}{2}\sum_{ij\alpha\beta\sigma} t^{\alpha\beta}_{ij}
 \hat{z}^\dagger_{i\alpha\sigma} \hat{z}_{j\beta\sigma} f^\dagger_{i\alpha\sigma} f_{j\beta\sigma}
 + \sum_{i\alpha\sigma}  (\epsilon_\alpha -\mu) f^\dagger_{i\alpha\sigma}
 f_{i\alpha\sigma}
 \nonumber\\
 & - \lambda_{i\alpha\sigma}[f^\dagger_{i\alpha\sigma}
 f_{i\alpha\sigma}-\frac{1}{2}(\hat{n}^a_{i\alpha\sigma}
 - \hat{n}^b_{i\alpha\sigma})] + H^S_{\mathrm{int}}.\nonumber
\end{align}
%\end{eqnarray}
Here, $\lambda_{i\alpha\sigma}$ is a Lagrange multiplier used
 to enforce the constraint in Eq.~\eqref{Eq:constraint-in-bosons}.
In addition,  $H^S_{\mathrm{int}}$ is the interaction Hamiltonian in Eq.~(2) of the main text rewritten in the slave-spin representation
$H_{\mathrm{int}}\rightarrow H_{\mathrm{int}}(\mathbf{S})$~[10], %\cite{SM_Yu_2012}
and
subsequently with the slave-spin operators
substituted by the Schwinger bosons.
The quasiparticle spectral weight
\begin{equation}\tag{S8}
\label{Eq:qpWeightZ}
Z_{i\alpha\sigma}=
|z_{i\alpha\sigma}|^2 \equiv
|\langle \hat{z}_{i\alpha\sigma}\rangle|^2 .
\end{equation}
A metallic phase corresponds to $Z_{i\alpha\sigma}>0$, and a Mott insulator corresponds to $Z_{i\alpha\sigma}=0$
in all orbitals with a gapless spinon spectrum.

After decomposing the boson and spinon operators and treating the constraint on average,
we obtain two saddle-point Hamiltonians for the spinons and the Schwinger bosons, respectively:
%\begin{eqnarray}
\begin{align}\tag{S9}
 \label{Eq:Hfmf}
 H^{\mathrm{mf}}_f &=  \sum_{k\alpha\beta}\left[ \xi^{\alpha\beta}_{k} \langle \tilde{z}^\dagger_\alpha \rangle
  \langle \tilde{z}_\beta \rangle + \delta_{\alpha\beta}(\epsilon_\alpha-\lambda_\alpha+\tilde{\mu}_\alpha-\mu)\right] f^\dagger_{k\alpha} f_{k\beta},\nonumber\\
 \nonumber\\
 \label{Eq:HSSmf}
 H^{\mathrm{mf}}_{S} &= \sum_{\alpha\beta} \left[Q^f_{\alpha\beta}
 \left(\langle \tilde{z}^\dagger_\alpha\rangle \tilde{z}_\beta+ \langle \tilde{z}_\beta\rangle \tilde{z}^\dagger_\alpha\right)
 + \delta_{\alpha\beta}\frac{\lambda_\alpha}{2} (\hat{n}^a_\alpha-\hat{n}^b_\alpha)\right] \nonumber\\
 &+ H^S_{\mathrm{int}},\tag{S10}
\end{align}
%\end{eqnarray}
where $\delta_{\alpha\beta}$ is Kronecker's delta function,
$\xi^{\alpha\beta}_{k}=\frac{1}{N}\sum_{ij\sigma} t^{\alpha\beta}_{ij} e^{ik(r_i-r_j)}$, and
%\begin{eqnarray}
\begin{align}\tag{S11}
\label{Eq:Qf}
Q^f_{\alpha\beta} &= \sum_{k\sigma}\xi^{\alpha\beta}_k\langle f^\dagger_{k\alpha\sigma}
f_{k\beta\sigma}\rangle/2,\\
\tag{S12}\label{Eq:tildez} \tilde{z}^\dagger_\alpha &= \langle P^+_\alpha\rangle a^\dagger_\alpha b_\alpha \langle P^-_\alpha\rangle.
\end{align}
%\end{eqnarray}
In addition,
$\tilde{\mu}_\alpha$ is an effective onsite potential defined as
\begin{equation}\tag{S13}
\label{Eq:tilde-mu}
\tilde{\mu}_\alpha = 2\bar{\xi}_\alpha \eta_\alpha
\end{equation}
where
\begin{equation}\tag{S14}
\label{Eq:tilde-bar-epsilon}
\bar{\xi}_\alpha = \sum_\beta\left( Q^f_{\alpha\beta} \langle\tilde{z}_\alpha^\dagger\rangle \langle \tilde{z}_\beta \rangle  + \rm{c.c.} \right)
\end{equation}
and
\begin{equation}\tag{S15}
\label{Eq:eta}
\eta_\alpha = (2n^f_\alpha-1)/[4n^f_\alpha(1-n^f_\alpha)],
\end{equation}
with $n^f_\alpha=\frac{1}{N}\sum_k \langle f^\dagger_{k\alpha} f_{k\alpha} \rangle$.

Eqs.~\eqref{Eq:Hfmf} and \eqref{Eq:HSSmf} represent the main formulation of the $U(1)$
slave-spin approach at the saddle-point level. Note that in this
approach,
the spinon dispersion
(along with the dispersion of the physical electrons) is naturally
 renormalized by the quasiparticle spectral weights $\sqrt{Z_\alpha Z_{\beta}}$ and the effective onsite potential $\tilde{\mu}_\alpha$.
We study the metal-to-insulator transitions in the paramagnetic phase preserving the translational symmetry.
The latter allows us to drop the spin and/or site indices
of the Schwinger bosons (slave spins) and the Lagrange multiplier $\lambda_\alpha$ in the above
saddle-point equations. We refer to
Refs.~[10] %\onlinecite{SM_Yu_2012}]
and [16] %\onlinecite{SM_Yu_2017}]
for a detailed derivation of these
saddle-point Hamiltonians. The
parameters
$z_\alpha$ and $\lambda_\alpha$
are solved self-consistently.

The focus of the present work is
the effect of the nematicity on the orbital selectivity. We have therefore treated only the diagonal part of
the Hund's coupling $J_{\rm{H}}$. The Mott and orbital-selective Mott transitions are related to the charge $U(1)$ symmetry of the system.
Therefore, retaining only the $U(1)$ instead of the $SU(2)$ spin symmetry does not qualitatively affect the nature of the transitions. But quantitative difference, such as the critical $U$ values for the transitions, may exist, as discussed in
DMFT studies.[51,52] %\cite{SM_Ishida_Liebsch_2010,SM_Imada_2010}
In the slave-spin calculation,
we reach similar conclusions:
for a fixed $J_{\rm{H}}/U$ ratio, the critical $U$ values for the transitions will increase when the off-diagonal terms are taken into account because they increase the ground state degeneracy~[12]. %\cite{SM_Yu2011}.
To incorporate this effect within our approach,
we take a larger $J_{\rm{H}}/U$ ratio in our slave-spin calculation.
In practice, we adopt $J_{\rm{H}}/U=0.25$ which is
 larger than the realistic value of  $J_{\rm{H}}/U\sim0.15$
~[31,53]. %\cite{SM_Yin_2011,SM_Das_2015}.
In the C4-symmetric case, this
 procedure has been verified to generate consistent results with ARPES for the case of Fe(Te,Se) substitution series~[54].%\cite{SM_Liu_2015}.

\subsection{The band splittings}
\begin{figure}[h!]
\centering\includegraphics[%scale=0.28
width=150mm
]{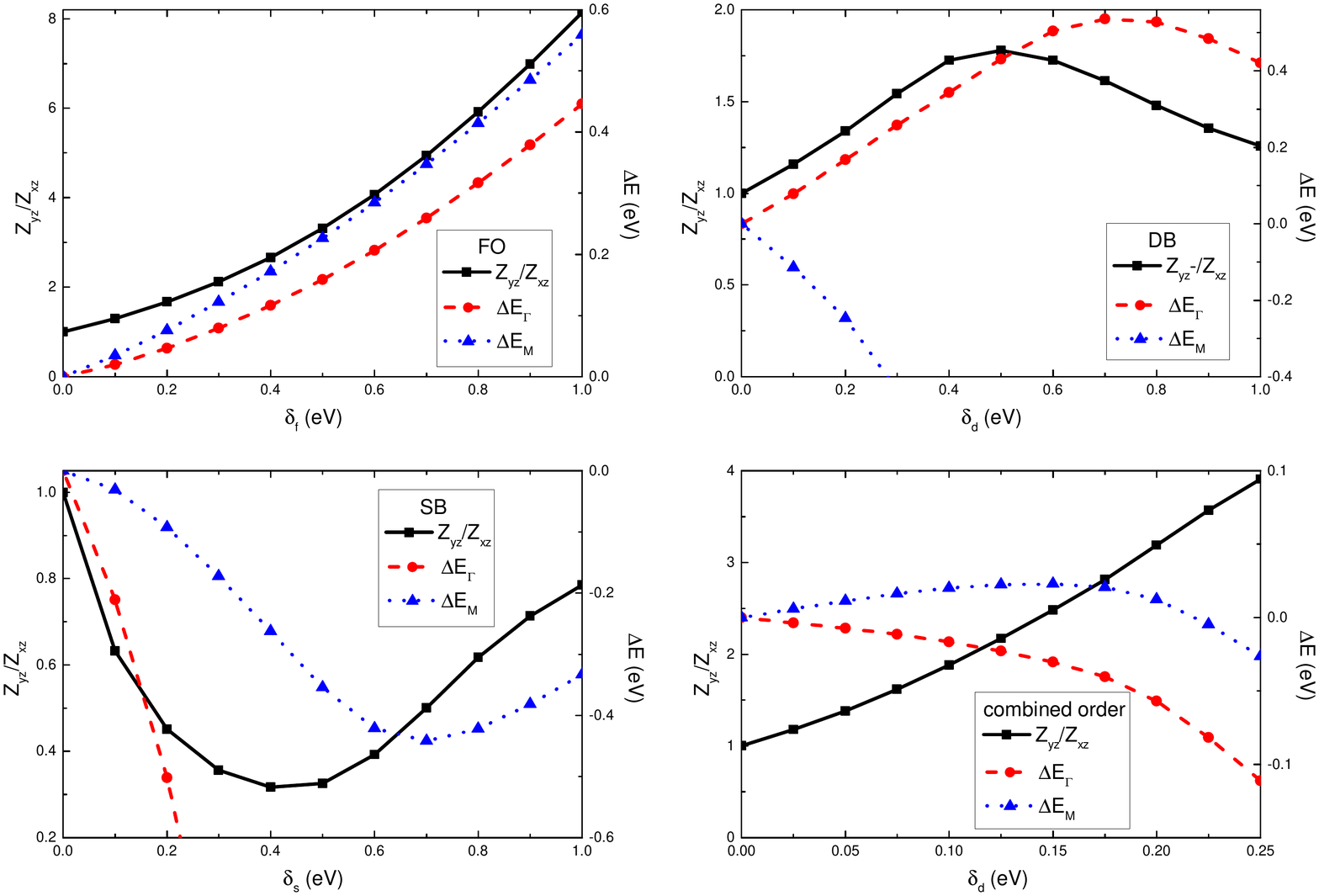}
\caption{(Color online) Evolution of $Z_{yz}/Z_{xz}$ (black solid) and band splittings at $\Gamma$ and M points, $\Delta E_\Gamma$ (red dashed) and $\Delta E_{\rm{M}}$ (blue dot), with various nematic orders at $J_{\rm{H}}/U=0.25$ and $U=3.5$ eV. FO, DB, and SB denote ferro-orbital order $\delta_f$, $d$-wave bond nematic order $\delta_d$, and $s$-wave nematic order, respectively. The combined order refers to the one with $\delta_f/4=\delta_d=\delta_s$.}
\label{Sfig:7}
\end{figure}

Here we give the explicit expressions of the non-interacting
band splittings at $\Gamma$ and M points.
From Eq.~(4) of the main text, we find the bare band splitting at $\Gamma$ is
\begin{equation}\tag{S16}
 \Delta E_\Gamma = E_{xz}(0,0)-E_{yz}(0,0) = 2(\delta_f-4\delta_s),
\end{equation}
and the band splitting at M point is
\begin{equation}\tag{S17}
 \Delta E_{\rm{M}} = E_{xz}(0,\pi)-E_{yz}(\pi,0) = 2(\delta_f-4\delta_d).
\end{equation}
It is easy to see that for $\delta_f/4=\delta_d=\delta_s$, the bare band splittings $\Delta E_\Gamma=\Delta E_{\rm{M}}=0$:
The effects of the different nematic order components cancel with each other,
in contrast to when either the ferro-orbital order $\delta_f$, or the bond nematic order $\delta_d$ or $\delta_s$ acts alone.
Taking into account the electron correlations, both $\Delta E_\Gamma$ and $\Delta E_{\rm{M}}$ will be renormalized from their bare values. But as shown
in Fig.~\ref{Sfig:7}, the cancelation effect is only prominent when taking a combined nematic order.
% Fig.~\ref{Sfig:7}(d) is similar to Fig 4 of the main text, but for $U=3.5$ eV.

\subsection{Supplemental figures}

\begin{figure}[h!]
\centering\includegraphics[%scale=0.28
width=120mm
]{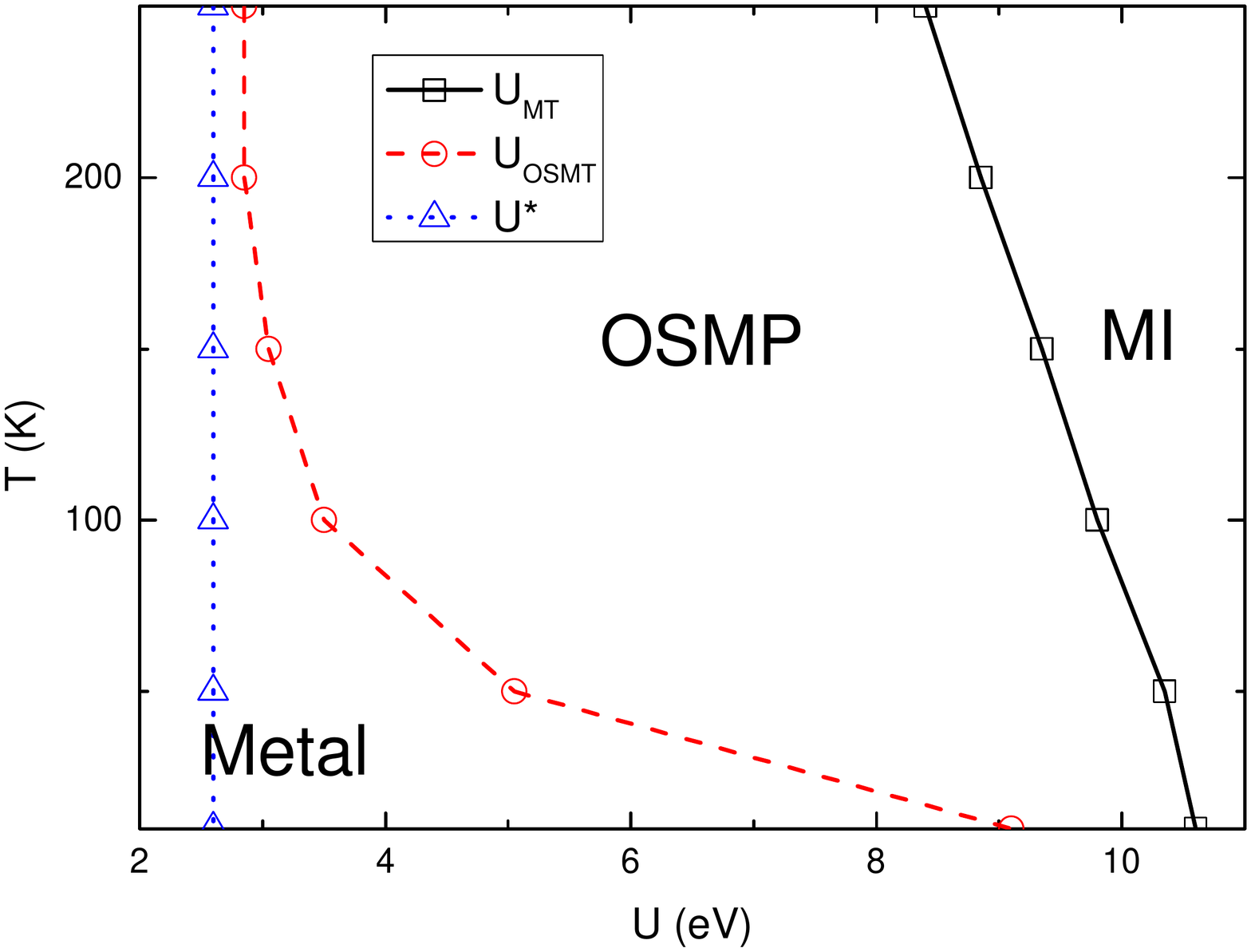}
\caption{(Color online) Finite-temperature phase diagram of the five-orbital Hubbard model for FeSe with tetragonal symmetry at $J_{\rm{H}}/U=0.25$. $U_{\rm{MT}}$ and $U_{\rm{OSMT}}$ are the
threshold
$U$ values for the
transition (T=0) or crossover ($T >0$) into a Mott insulator or an orbital-selective Mott phase,
respectively. $U^\star$ refers to the crossover between the weakly correlated metal to the strongly correlated metal.}
\label{Sfig:3}
\end{figure}

\begin{figure}[h!]
\centering\includegraphics[%scale=0.28
width=120mm
]{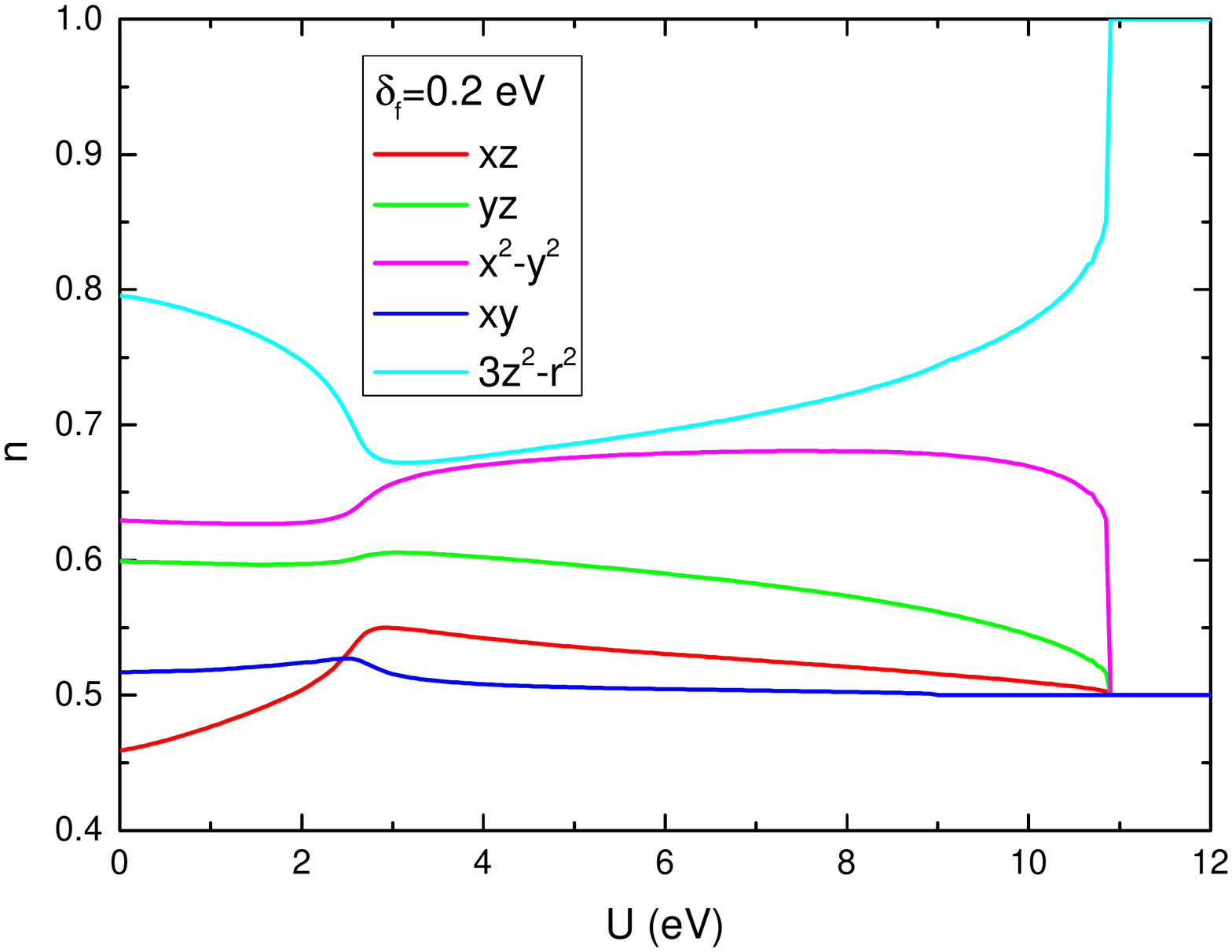}
\caption{(Color online) Electron density per orbital per spin flavor at
$J_{\rm{H}}/U=0.25$
and with a bare ferro-orbital order $\delta_f=0.2$ eV.}
\label{Sfig:4}
\end{figure}

\begin{figure}[h!]
\centering\includegraphics[%scale=0.28
width=120mm
]{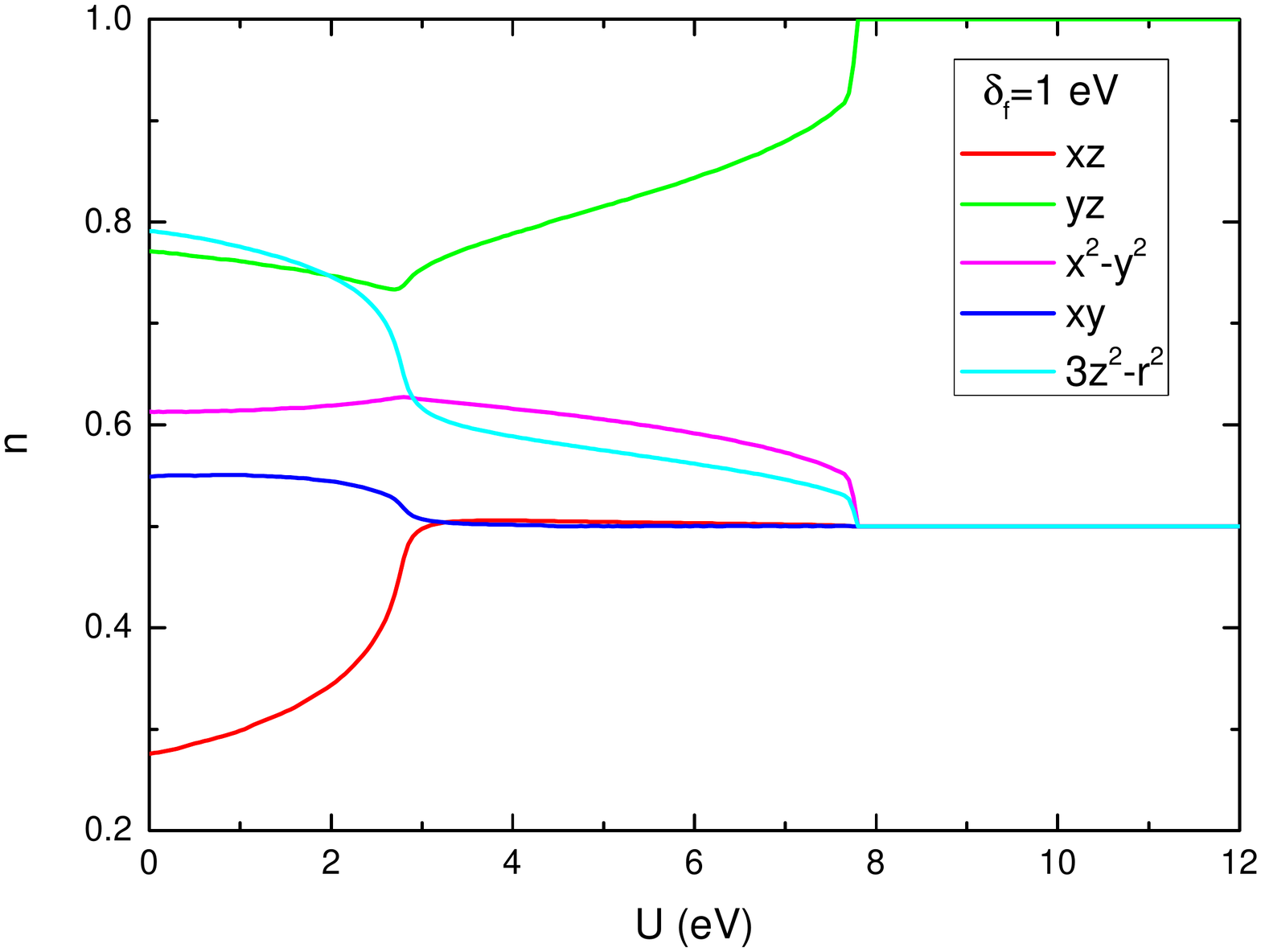}
\caption{(Color online) Electron density per orbital per spin flavor at
$J_{\rm{H}}/U=0.25$
and with a bare ferro-orbital order $\delta_f=1$ eV.}
\label{Sfig:5}
\end{figure}

\begin{figure}[h!]
\centering\includegraphics[%scale=0.28
width=100mm
]{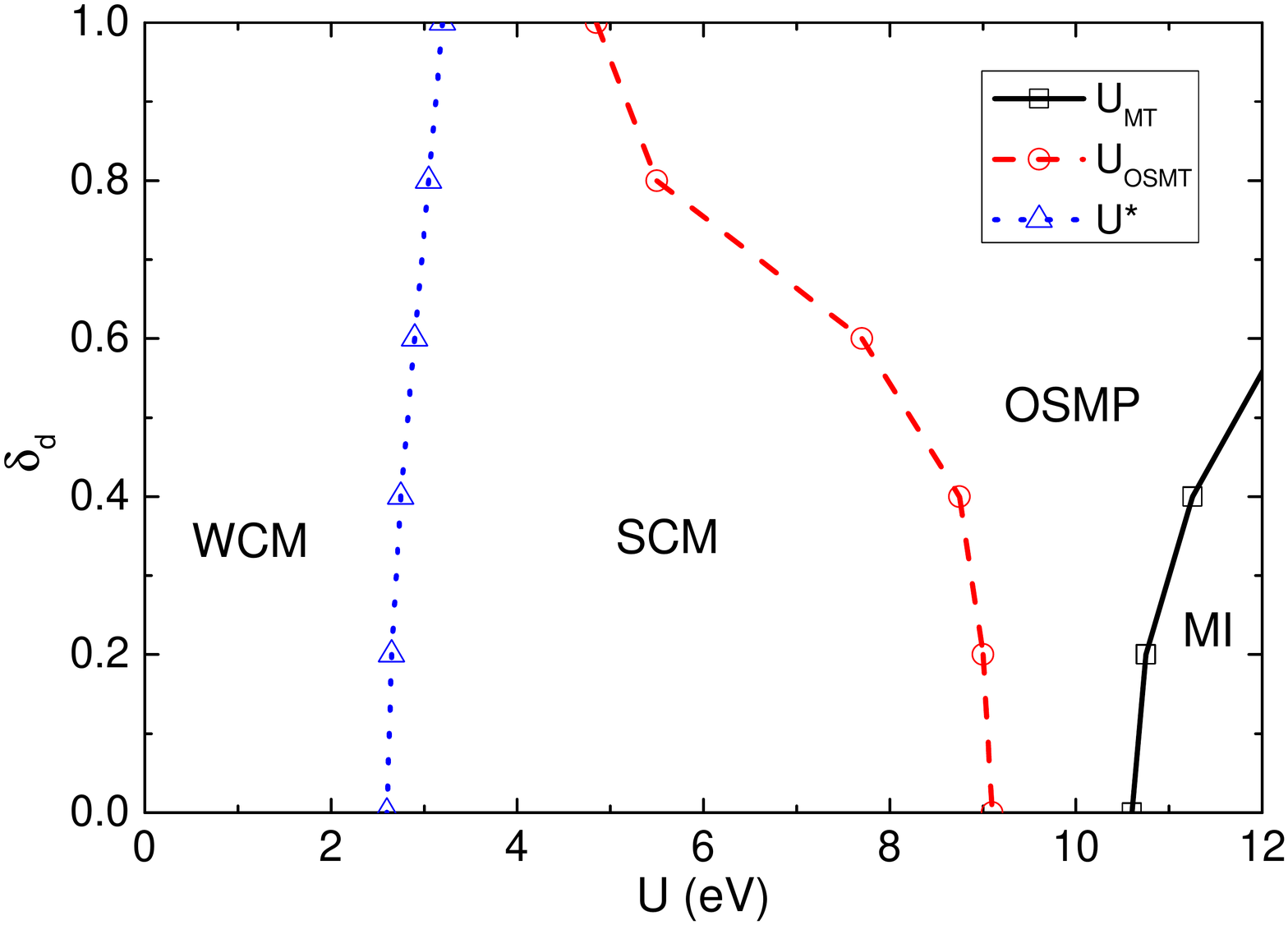}
\caption{(Color online) Ground-state phase diagram of the five-
orbital Hubbard model for FeSe with a $d$-wave nearest-neighbor nematic order $\delta_d$ at $J_{\rm{H}}/U=0.25$.}
\label{Sfig:6}
\end{figure}

\end{document}